\documentclass[preprint2,longabstract]{aastex}

\renewcommand{\b}[1]{\mbox{\boldmath $#1$}}

\def\cal#1{{\cal #1}}

\def\m@th{\mathsurround=0pt}
\def\n@space{\nulldelimiterspace=0pt \m@th}
\def\biggg#1{{\mbox{$\left#1\vbox to 20.5pt{}\right.\n@space$}}}

\def\beginenum{\begin{enumerate}}
\def\endenum{\end{enumerate}}
\def\bitem{\begin{itemize}}
\def\eitem{\end{itemize}}
\def\bray{\begin{array}}
\def\eray{\end{array}}
\def\begindoc{\begin{document}}
\def\enddoc{\end{document}}
\def\bq{\begin{equation}}
\def\eq{\end{equation}}
\def\bqy{\begin{eqnarray}}
\def\eqy{\end{eqnarray}}
\def\bqyn{\begin{eqnarray*}}
\def\eqyn{\end{eqnarray*}}
\def\bc{\begin{center}}
\def\ec{\end{center}}
\def\bfll{\begin{flushleft}}
\def\efll{\end{flushleft}}
\def\bflr{\begin{flushright}}
\def\eflr{\end{flushright}}
\newcommand{\Avec}{\mbox{\boldmath $A$}}
\newcommand{\Bvec}{\mbox{\boldmath $B$}}
\newcommand{\Evec}{\mbox{\boldmath $E$}}
\newcommand{\Fvec}{\mbox{\boldmath $F$}}
\newcommand{\Gvec}{\mbox{\boldmath $G$}}
\newcommand{\Rvec}{\mbox{\boldmath $R$}}
\newcommand{\Uvec}{\mbox{\boldmath $U$}}
\newcommand{\Vvec}{\mbox{\boldmath $V$}}
\newcommand{\evec}{\mbox{\boldmath $e$}}
\newcommand{\jvec}{\mbox{\boldmath $j$}}
\newcommand{\kvec}{\mbox{\boldmath $k$}}
\newcommand{\nvec}{\mbox{\boldmath $n$}}
\newcommand{\uvec}{\mbox{\boldmath $u$}}
\newcommand{\vvec}{\mbox{\boldmath $v$}}
\newcommand{\wvec}{\mbox{\boldmath $w$}}
\newcommand{\xvec}{\mbox{\boldmath $x$}}
\newcommand{\omegavec}{\mbox{\boldmath $\omega$}}
\newcommand{\Omegavec}{\mbox{\boldmath $\Omega$}}

\usepackage{spr-astr-addons}    

\usepackage{color}

\RequirePackage{color}
\def\imagei{\centerline{\color[gray]{.75}\rule{\hsize}{4pc}}}%
\def\imageii{\centerline{\color[gray]{.75}\rule{4pc}{4pc}}}%

\newcommand{\vdag}{(v)^\dagger}
\newcommand{\emaila}{authors@email.com}

\begin{document}
%
\title{Dynamical fast flow generation/acceleration in dense degenerate two-fluid
plasmas of astrophysical objects}

\shorttitle{<Fast flow generation in degenerate astrophysical objects>}
\shortauthors{<Kotorashvili et al >}

\author{ K. Kotorashvili\altaffilmark{1}}
\altaffiltext{1}{Department of Physics, Faculty of Exact \&
Natural Sciences, Javakhishvili Tbilisi State University,
Tbilisi 0179, Georgia}
\author{ N. Revazashvili\altaffilmark{1}}
\altaffiltext{1}{Department of Physics, Faculty of Exact \&
Natural Sciences, Javakhishvili Tbilisi State University,
Tbilisi 0179, Georgia}
\author{ N.L. Shatashvili\altaffilmark{1,2}}
\altaffiltext{1}{Department of Physics, Faculty of Exact \&
Natural Sciences, Javakhishvili Tbilisi State University,
Tbilisi 0179, Georgia} \altaffiltext{2}{Andronikashvili Institute
of Physics, TSU, Tbilisi 0177, Georgia}

\begin{abstract}
We have shown the generation/amplification of fast macro-scale
plasma flows in the degenerate two-fluid astrophysical systems
with initial turbulent (micro--scale) magnetic/velocity fields
due to the Unified Reverse Dynamo/Dynamo mechanism. This process
is simultaneous with and complementary to the micro-scale unified
dynamo. It is found that the generation of macro--scale
flows is an essential consequence of the magneto-fluid coupling;
the generation of macro--scale fast flows and magnetic
fields are simultaneous, they grow proportionately.
The resulting dynamical flow acceleration
is directly proportional to the initial turbulent magnetic
(kinetic/magnetic) energy in degenerate e-i (degenerate e-p)
astrophysical plasma; the process is very sensitive to
both the degeneracy level of the system and the
magneto-fluid coupling. In case of degenerate e-p plasma,
for realistic physical parameters, there always exists
such a real solution of dispersion relation for which the formation
of strong macro-scale flow/outflow is guaranteed;
the generated/accelerated locally super-Alfv\'enic flows
are extremely fast with Alfv\'en Mach number $> 10^3$
as observed in a variety of astrophysical outflows.

\end{abstract}


\maketitle


\section{Introduction}

Several recent studies were devoted to the mechanisms that
explain flow/outflow formations in stellar atmospheres.
Flows as well as transient jets are observed in solar atmosphere --
their role in the dynamics and heating of multi-scale
complex-structure solar corona is already well appreciated.
Flows are found crucial in astrophysical disks [see e.g.
\citep{vinod,zanni}, \citep{SY-DJ},\\
\citep{bodo} and references therein] and their corona,
in inter- and extra-galactic environments.
More extended large scale outflows are met in various
astrophysical settings, e.g. AGN relativistic jets,
protostellar jets, being the collimated long-lived
structures related to accreting disks surrounding the
compact objects [see e.g. \citep{Begelman} and
references therein]. In this view, in addition
to the study of star evolution dynamics,
it is important to uncover the contribution (if any)
of flow dynamics in compact objects outer layers
to the formation of large-scale jets/outflows.

Study of the multi-scale dynamics of compact object's multi-component
magnetospheres attracted the interest to solve phenomena
related to star evolution problem. Among these investigations the
studies on equilibrium structure formations based on
so called Beltrami-Bernoulli (BB) class of equilibria model
\citep{Shiraishi,iqbal-3,pino,Multi-B}
opened the new channels for exploring the heating of atmospheres
as well as the problems of large-scale magnetic and/or velocity
field generation \citep{mmns-1,ymois}; \cite{osym};
\citep{mnsy},\\
\citep{msms}; protostellar disk-jet structure formation \citep{yso}.
The examination of BB states were also performed
for highly dense and degenerate plasmas applicable
to compact star conditions (mean inter-particle distance is
smaller than the de Broglie thermal wavelength so that
particle energy distribution was dictated by Fermi-Dirac
statistics) [see \citep{degenerate} and references therein].
Such highly dense/degenerate plasmas are also found in various
astrophysical / cosmological environments, in
laboratories devoted to inertial confinement, in high energy
density physics \citep{Dunne-1,Dunne-2,Yanovski,Dunne-3}.

The density (determinant of degeneracy level) varies over
many orders of magnitude in astro-settings.
Compact astrophysical objects like white and brown dwarfs,
neutron stars, magnetars with characteristic electron number
densities within $(10^{26}--10^{32})\,cm^{-3}$ are the
natural habitats for degenerate matter \\
\citep{Chandra1,Chandra2};\\
\citep{Compact-1,Begelman};\\
\citep{Compact-2,White};\\
\citep{Sturok-3,Beloborodov};\\
\citep{Shukla-1,Shukla-2}. \\
Rest frame e-p density near pulsar surface $\geq 10^{11}\,cm^{-3}$
\citep{Misha}, while in the MeV epoch of the
early Universe, it  can be  $10^{32}\,cm^{-3}$
\citep{Weinberg}. Intense e-p pair creation takes place during the
gravitational collapse of massive stars \citep{Stenflo}
with estimated density $\sim 10^{34}\,cm^{-3}$ \ \citep{ruffini}.
In GRB sources \citep{aksenov} there may exist a
superdense e-p plasma with density $(10^{30}-10^{37})\,cm^{-3}$.
The consequences of degeneracy in a multi-component
plasma was extensively studied recently in terms of multi-scale behavior
accessible to such systems \citep{BSM_deg,SMB_multi}, \\
\citep{BS-flow} to explore its role in
the dynamics of star collapse while contraction
of its atmosphere; to predict various
phenomena in pre-compact era, or the compact
objects' dynamics since cooling process seems to be sensitive
to outer layers/atmosphere composition, structure and
their conditions.

Up to now there doesn't exist a precise model of
atmospheres of White Dwarfs (WDs) although recent studies show
that a significant fraction of White Dwarfs are found to be
magnetic. Massive and cool White Dwarfs are found with high ($> 1\,KG$) fields
detected [see \citep[and references therein]{winget,kepler}].
Interestingly, the electron degeneracy manifests, explicitly,
only through the Bernoulli condition for the case of
mildly degenerate e-i plasma \citep{BSM_deg},
and as a result, such gas can sustain a qualitatively new state:
a nontrivial Double BB equilibrium at zero temperature.
It is extremely interesting how this effect will manifest or define
the fate of Unified Dynamo/RD process \citep{msms,lingam},
specifically in view of flow generation in the vicinity of
compact object, in several other astrophysical setups.

In \citep{BS-flow} the fast flow generation due to
magneto-fluid coupling (through catastrophe) near
the surface of dense degenerate e-i stellar atmospheres
was suggested finding that distance over which
acceleration appears is determined by the strength
of gravity and degeneracy parameter. Application of this
mechanism for White Dwarfs' atmospheres was examined
showing the possibility of the super-Alfv\'enic flow
generation for various surface parameters of WDs;
the simultaneous possibility of flow acceleration
and magnetic field amplification for specific
boundary conditions was explored in which the
degeneracy has a striking effect. For the understanding
of origin and evolution of dense compact objects;
for their cooling and accretion dynamics; to know
the magnetic fields dynamics/fate the inclusion of
time-dependency may become determining and crucial --
the unified Dynamo / Reverse Dynamo treatment \citep{msms,lingam}
can lead to additional significant effects on the formation of
large-scale flows and/or magnetic fields in astrophysical
objects with degenerate plasmas (e.g. outer layers of compact
objects).

The goal of present study is to explore the role of
degeneracy in the dynamical fast flow formation
for degenerate plasmas of astrophysical objects;
we will follow the methodology of Reverse Dynamo (RD)
mechanism proposed in (Mahajan et al. 2005) (demonstrating
that dynamo and RD operate simultaneously).
RD -- permanent dynamical feeding of flow kinetic
energy through an interaction of microscopic magnetic field
structures with weak flows was shown to be universal
property (indicating application for Solar atmosphere).
\citep{lingam} conjectured that an efficient RD may
be the source of observed astrophysical outflows
with Alfv\'en Mach number $\gg 1$. Due to the existence
of an intrinsic micro--scale in HMHD at which ion /
degenerate e-p kinetic inertia effects become important
it is possible to characterize long and short scales in a well
defined way (the macroscopic scale of the system is
generally much larger than the charged fluid skin depth).
We will examine the possible role played by Unified
Dynamo/RD mechanism in explaining the existence of large-scale
velocity and magnetic fields in degenerate two-fluid
plasmas of astrophysical objects. We will find the applications
for: 1) WDs with degenerate electrons and classical ions
assuming the density variations to be slow below the
catastrophe heights \citep{BS-flow}; 2) astrophysical objects
with degenerate e-p plasma for which the degeneracy effects
become crucial when the inertia of bulk e-p components
makes the effective skin depths much larger than
the standard skin depth \citep{SMB_multi}.

\section{Model Equations for Unified Reverse Dynamo / Dynamo mechanism for WDs}

In this section we start from the outer layers of compact objects,
specifically the case of WDs' -- end product of the star accretion
evolution -- that are considered to be stellar remnants featuring
global magnetic structures with field strengths within $1\,kG \div
1000\,MG$ \citep{hmfwd,Kawka}. Most of these objects are
higher-field magnetic WDs; a distribution of magnetic field
strengths appears to peak around $B>20\,MG$ \citep{schmidt,kulebi}.
In \citep{DAZ} it was shown that WD stars with developed
convective zones show stronger magnetic fields than hotter
stars; the mean mass of magnetic stars is on average larger than
the mean mass of nonmagnetic WD stars.
The effective temperatures of convective hydrogen-line
(DA) white dwarfs are in the range $(6000 - 15000)\,K$
and convective velocities are of the order of $\sim 1\,km/s$
at the base of the convection zone reaching maximum value $6\,km/s$
\citep[and references therein]{tremblay};
while for cool, magnetic, polluted hydrogen atmosphere WDs (DAs)
it was found to be $(19.8 \pm 1.7)\,km/s $
\citep[and references therein]{DAZ}. Many WDs have
much stronger ($\gtrsim 3$\,MG) surface magnetic fields;
this could be partially explained by the core
dynamo-generated fields \citep{ferrario}.
Even stronger ($\gtrsim 10^6$\,G) magnetic fields
could be confined within the WD's interior and not detectable
at the surface even as they cool \citep{cumming}.
Then, it is expected that the dynamical evolution of WD's convective
envelope / outer layers may define the final structure of its
interior as well as of atmosphere \citep{BS-flow}.

The simplified HMHD of \citep{BSM_deg} for a two-species system
of non-degenerate non relativistic ions, and degenerate
relativistic electrons embedded in a magnetic field --
a minimal model
that contains two disparate interacting scales --
can be useful for studying the Unified RD/Dynamo in WD's outer layers.
In this model the ion (${\b v}$) and degenerate electron
(${\b v}_e = {\b v}- {\bf j}/eN$) flow velocities are different
even in the limit of zero electron inertia.
In its dimensionless form, HMHD equations for degenerate electron-ion
plasma reduce to:
\begin{equation}
\frac{\partial \boldsymbol{b}}{\partial t} = \nabla \times
\left[\left(\boldsymbol{v} - \frac{\alpha}{N}\left(\nabla
\times \boldsymbol{b}\right)\right)\times \boldsymbol{b}\right] \ ,
\label{WD1}
\end{equation}
\[
\frac{\partial \boldsymbol{v}}{\partial t} = \frac{1}{N}
\left(\nabla \boldsymbol{\times }\boldsymbol{b}\right)\boldsymbol{\times }
\boldsymbol{b} + \boldsymbol{v}\boldsymbol{\times }\nabla \boldsymbol{\times }
\boldsymbol{v}
\]
\begin{equation}
\qquad - \nabla \left({\beta }_0{ln N\ }-{\mu }_0\left(G_d\gamma \right)
+ \frac{{\boldsymbol{v}}^{\boldsymbol{2}}}{2} - \frac{R_A}{R}\right) \ .
\label{WD2}
\end{equation}
where \ ${\boldsymbol{b}}=e{\bf B}/m_ic$ \ and it was assumed, that
electron and proton laboratory-frame densities are nearly equal -
$N_e\simeq N_i=N$ \ [rest-frame density \ $n =
N/\gamma ({\b v})$ \ with \ $\gamma ({\b v})\simeq
\gamma_e$ \ being a Lorentz factor for electrons];
the density is normalized to $N_0$ (the
corresponding rest-frame density is \ $n_0$); the magnetic
field is normalized to some ambient $B_0$ ;
all velocities are measured in terms of the corresponding Alfv\'en
speed \ $V_A=B_0/\sqrt{4\pi N_0m_i}$ ; all lengths are
normalized to the characteristic length-scale of the system, WD-radius
$R_w $. $\beta_0$ is an equilibrium plasma beta;
\ $\mu_0=m_ec^2/m_iV_A^2$ ; \
$R$ is the radial distance from the center of WD normalized to
its radius \ $R_{W}\ [\sim (0.008 - 0.02)\ R_{\odot}]$ and
$R_A=GM_{W}/R_{W}V_A^2$ ($G$ -- gravitational constant,
$M_W$ -- WD mass);  dimensionless parameter $\alpha =
\lambda_i/R_w$ . Here the ion-skin-depth
$\lambda_i =c/\omega_{pi} \sim (10^{-5}-10^{-7})\,cm$ for a
typical cold magnetic WDs with degenerate electron densities
$\sim (10^{25}-10^{29})\,cm^{-3}$ and magnetic fields $\sim
(10^5-10^{9})\,G $, temperatures \ $\sim (40000-6000)\,K $.
The degeneracy induced effective mass factor for strongly
degenerate electron plasma is determined by the plasma rest frame density,
$G_{d}=[ 1+(n/n_{c})^{2/3}]^{1/2}$ \ for arbitrary \ $n/n_c$,
\ with \ $n_c = 5.9\times 10^{29}cm^{-3}$
being the critical number-density.
Comparing the terms in total pressure on the r.h.s.
of (\ref{WD2}) for above parameters, one can see the dominance
of electron fluid degeneracy pressure.

Following the standard procedure (Mahajan et al. 2005))
let's assume that our total fields are composed of some
ambient seed fields as well as density and fluctuations about
them with account of degeneracy effects:
\begin{equation}
n = n_0 + \delta n \ ; \quad \boldsymbol{b}={\boldsymbol{b}}_{\boldsymbol{0}}
+ \boldsymbol{H}+\tilde{\boldsymbol{b}} \ , \quad
\boldsymbol{v} = {\boldsymbol{v}}_{\boldsymbol{0}}
+ \boldsymbol{U} + \tilde{\boldsymbol{v}} \ ,
\label{WD3}
\end{equation}
where $n_0=const, \ {\boldsymbol{b}}_{\boldsymbol{0}}$ , \
${\boldsymbol{v}}_{\boldsymbol{0}}$ are equilibrium density and
the equilibrium fields; $\boldsymbol{H}$ , $\boldsymbol{U}$ ,
$\delta n$ are the macroscopic fluctuations; and $\tilde{\boldsymbol{b}}$,
$\tilde{\boldsymbol{v}}$ are the microscopic fluctuations, respectively;
we have ignored the microscopic density fluctuation due to its
higher order contribution;
also the wave coupling is beyond the scope of this study. We emphasize
here, that the energy reservoir comes from the background fields
that may have both macro-scale and micro-scale components.
Their energy feeds the macro- and micro-scale fluctuations of
density and fields. It is natural to assume that in HMHD
the equilibrium fields are the solutions of so called Double
Beltrami Equations \citep{MY,mmns-1,msms}:
\begin{equation}
\boldsymbol{b}_0 = a \left(\boldsymbol{v}_0-\alpha\,
\nabla \times \boldsymbol{b}_0\right)\ , \ \ \
\boldsymbol{b}_0+\alpha\,\nabla \times \boldsymbol{v}_0
= d \boldsymbol{v}_0\boldsymbol \ ,
\label{WD4}
\end{equation}
together with the Bernoulli Condition
(at \ ${\beta }_0{ln {N}\ }\ll {\mu }_0\,G_d(N)
\gamma \ , \ \gamma ({\b v})\sim 1$ \ for our problem of interest):
\begin{equation}
\nabla \left({\mu }_0G_d(\delta n)+
\frac{{{\boldsymbol{v}}_{\boldsymbol{0}}}^{\boldsymbol{2}}}{2}
-\frac{R_A}{R}\right)=0 \ ,
\label{WD5}
\end{equation}
where \ $a$ \ and \ $d$ \ are dimensionless constants related
to the two invariants: the magnetic helicity \ $h_1=\int{({\bf A}
\cdot {\bf b})\,d^3x}$ \ and the generalized helicity \
$h_2=\int{({\bf A}+{\boldsymbol{v}})\cdot ({\bf b}+\nabla
\times \boldsymbol{v}) \,d^3x}$ \ of the system with \ ${\bf
A}$ \ being the dimensionless vector potential.
Notice, that in this approximation the electron
vorticity is primarily magnetic \ (${\boldsymbol{b}}_0$)
\ while the ion vorticity  \ (${\b b}_0+\nabla \times {\b v}_0$)
has both kinematic and magnetic parts. In \citep{BS-flow} it
was shown that due to the degenerate density both the
velocity and magnetic fields undergo catastrophe at
some height from the WD's surface. In present paper
we assume that the distances are below this point so that
macro-scale fluctuations of density and fields are slowly
varying functions. Also, for simplicity we assume
that our zeroth-order fields are wholly
at the microscopic scale. This allows us to create a hierarchy
in the micro--fields; the ambient fields are much greater than
the fluctuations at the same scale \ ($|\tilde{\b b}|\ll |{\b b}_0 |,
|\tilde{\b v}|\ll |{\b v}_0| $). This closure model accounts
properly the self-consistent feedback of the micro--scale
in the evolution of both macro-scale fields ${\bf H}$ and
${\bf U}$, as well as the role of the Hall current
(especially in the dynamics of the micro--scale)
[see Mininni et al. 2003, Mahajan et al. 2005 for details].

The invariant helicities control the final results through the
Beltrami scales $a$ and $d$. We choose these constants so that
the characteristic scales [inverse of \ ${\frac{1}{2}}[(d-a^{-1})
\pm ((d+a^{-1})^2 -4)^{-1/2}]$ ] become vastly separated
\citep{MY,mmns-1}. In the astrophysically relevant regime
of disparate scales (the size of the structure is much
greater than the ion skin depth), we deal with two extreme
cases (in the analysis below we use $\lambda $ for the
micro-scale and $\mu $ for the macro-scale): (1) $a \sim d \gg 1$
and ($(a-d)/ad \ll 1$ [$\lambda \sim d$ and $\mu \sim (a-d)/ad $  ],
and (2) $a\sim d \ll 1$ and ($a-d)/ad \gg 1$ [$\lambda
\sim a-a^{-1}$ and $\mu \sim d-a$].

Astrophysical objects, including compact stars like WD
are macro-scale, then, consistent with the main objectives
of this paper, we assume that basic reservoir
[from which system generates macro-scale fields] is at
micro-scale; neglecting macro-scale equilibrium
component we find that the velocity and magnetic fields get
linearly related as \ ${\boldsymbol{v}}_0 = (\lambda + a^{-1})
{\boldsymbol{b}}_0$  \ yielding \ ${\boldsymbol{v}}_{0e} =
{\boldsymbol{v}}_0 - \alpha\nabla \times
{\boldsymbol{b}}_0 = {(\lambda +a^{-1})\boldsymbol{b}}_0
-\lambda{\boldsymbol{b}}_0 = {a^{-1}\boldsymbol{b}}_0$ \
and leading to
\begin{equation}
\dot{\tilde{\boldsymbol{b}}}=({a^{-1}}{\boldsymbol{H}}
- {\boldsymbol{U}})\cdot \nabla {\boldsymbol{b}}_0 \ ,
\label{WD6}
\end{equation}
\begin{equation}
\dot{\tilde{\boldsymbol{v}}} = ({\boldsymbol{H}} -(\lambda +a^{-1})
{\boldsymbol{U})\cdot \nabla {\boldsymbol{b}}_0} \ .
\label{WD7}
\end{equation}
Using (\ref{WD6},\ref{WD7}) we obtain for macro-scale fields following:
\begin{equation}
\ddot{\boldsymbol{H}}=r\,\nabla \times \boldsymbol{H} \ , \qquad
\ddot{\boldsymbol{U}}=-\ \nabla \times (s\,\boldsymbol{U}-q\,\boldsymbol{H})\ ,
\label{WD10}
\end{equation}
where the constants $q$, $r$ and $s$ :
\[
q = \lambda^2 \frac{b_0^2}{6} \ , \qquad r =
- \lambda \frac{ b_0^2}{3}(1-\lambda{a^{-1}}-{a^{-2}}) \ ,
\]
\begin{equation}
s = \lambda \frac{b_0^2}{6}\left[{({\lambda +a^{-1}}})^2-1\right]
\label{WD11}
\end{equation}
are determined by DB parameters $a$ and $d$ (hence, by the ambient magnetic and
generalized helicities) and scales directly with the ambient turbulent
energy $\sim {\boldsymbol{b}}_0^2$ (${\boldsymbol{v}_0^2}$).
Performing Fourier analysis, we obtain:
\begin{equation}
\boldsymbol{U}=\frac{q}{s+r}\boldsymbol{H} \ .
\label{WD12}
\end{equation}
From Equations ({\ref{WD10}}, {\ref{WD11}}, {\ref{WD12}})
we observe that, to leading order, like in classical case \citep{msms},
${\boldsymbol{H}}$ evolves independently of ${\boldsymbol{U}}$,
but the reverse is not true: the evolution of ${\boldsymbol{U}}$
does require knowledge of ${\boldsymbol{H}}$. Hence, a choice of
Beltrami scales $a$ and $d$ that now reflect the helicities of
degenerate e-i system, fixes the relative amounts in ambient fields'
microscopic energy and, consequently, in the generated macroscopic fields
that grow proportionately to each other. In the subsection below
we show that the Unified RD/D mechanism affects the evolution
picture of outer layers of magnetic White Dwarfs (WD) predicted in
\citep{BSM_deg,SMB_multi} where it was shown that
when the star contracts its outer layer keeps the
multi-structure character although density in structures
becomes defined by electron degeneracy pressure.

\subsection{Reverse Dynamo for WDs' degenerate electron-ion plasma}

We can examine the two observationally justified extreme
cases for Beltrami scales:


\begin{figure}
\begin{center}
\includegraphics[scale=0.4,angle=0]{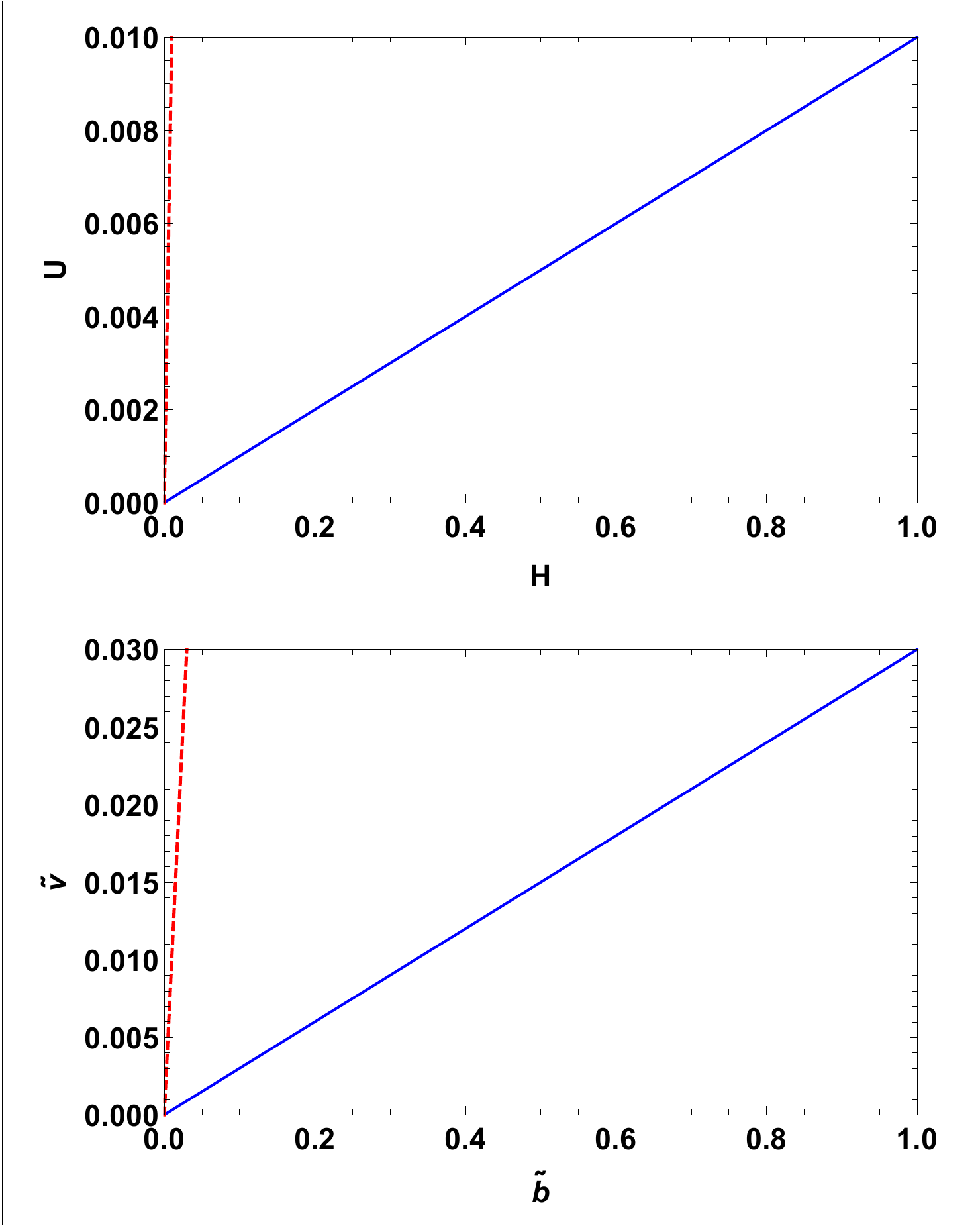}
\caption{Evolution of generated velocity and magnetic fluctuations (blue lines) for
Beltrami Parameneters $a\sim d =100$: top panel shows the macro-scale
velocity field ${\boldsymbol{U}}$ versus macro-scale magnetic field
${\boldsymbol{H}}$ while the bottom panel shows the micro-scale
${\boldsymbol{\tilde{v}}}$ versus micro-scale ${\boldsymbol{\tilde{b}}}$. Starting from
strongly super-Alfv\'enic ambient flow system arrives to sub-A;fv\'enic
macro-scale and micro-scale fluctuations, locally. Straight Dynamo
for both scales. Dotted lines in red show the Alffv\'enic flow
for both scales, respectively.  }
\label{Fig.1.}
\end{center}
\end{figure}


\begin{figure}
\begin{center}
\includegraphics[scale=0.4,angle=0]{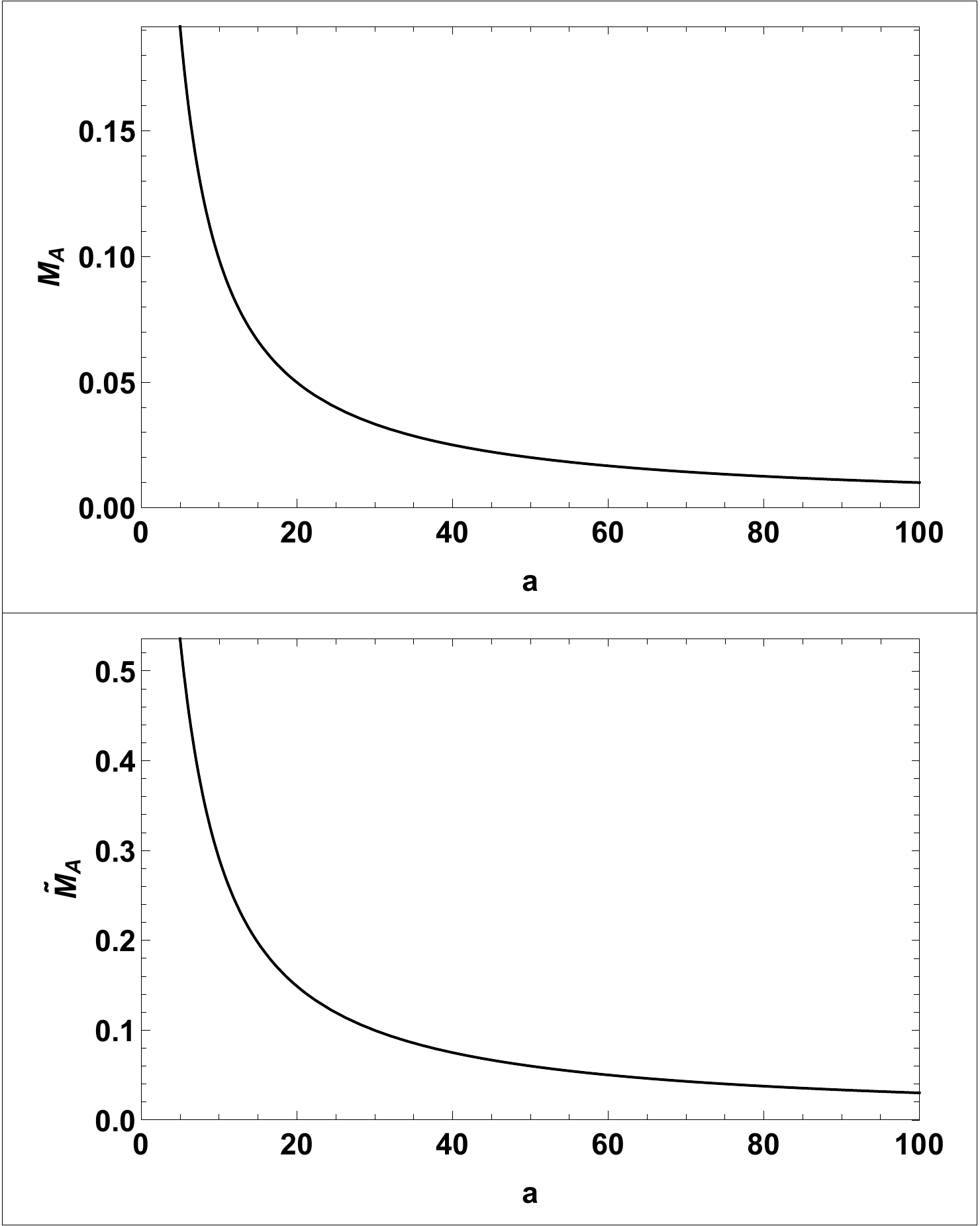}
\caption{Plot for Alfv\'en Mach Number versus $a > 1$ for macro-scale vector-fields
$M_A$ (top) and micro-scale vector-fields
${\tilde{M}}_A$ (bottom), respectively  for generated
velocity and magnetic fluctuations. Bigger the Beltrami scale $a$
smaller is the Alfv\'en Mach number that is $\ll 1$ for
all $a$-s. For both scales the Straight Dynamo scenario works
showing that Strong Magnetic fields (both macro-- and micro-- scales)
are generated from primarily kinetic micro-scale ambient state. }
\label{Fig.2.}
\end{center}
\end{figure}

(1) Example of primarily kinetic ambient fields:
$a\sim d\gg 1 \ , \ (\lambda \sim a \gg 1)$ implying
${\boldsymbol{v}}_0=(\lambda +a^{-1}){\boldsymbol{b}}_0\sim a\,
\boldsymbol{b}_0 \gg \boldsymbol{b}_0$. Such conditions
may be met in WD's photospheres, where the turbulent
velocity field at some stage can be dominant, although
some $\boldsymbol{b}_0$ is present as well. For these parameters,
the generated macro-fields have precisely the opposite ordering,
$\boldsymbol{U}\sim a^{-1}\boldsymbol{H}\ll \boldsymbol{H}$.
In this junction it is interesting to recall that
recent studies show that a significant fraction of
WDs are found to have rather strong surface magnetic fields
(see \citep{kepler,hollands,BS-flow} and references therein).
One of the evolution channels could be the amplification of
a seed field by a convective dynamo in the core–-envelope
boundary of the evolved progenitors (Ruderman \& Sutherland
1973; Kissin \& Thompson 2015).
It is, therefore, very important to show that the effects of
magneto-fluid couplings in outer layers of accreting stars may lead to
the dynamical evolution of their convective envelopes and generate
macro-scale magnetic field through macro-scale Dynamo mechanism.
Our analysis show that this process is maintained through
the generation of micro-scale fluctuations
${\boldsymbol{\tilde{v}}}$ and ${\boldsymbol{\tilde{b}}}$. In Fig.1
the relevant plots are presented. Fig.2 gives the results for so called Alfv\'en
Mach Number versus Beltrami scale $a > 1$ for macro-scale vector-fields
($M_A$) and micro-scale vector-fields (${\tilde{M}}_A$), respectively.
We see, that starting from primarily super-Alfv\'enic micro-scale
flows the magneto-fluid coupling guarantees the straight
Dynamo scenario -- the generated fields are sub-Alfv\'enic in both scales.
Created macro(micro)--scale flows and Magnetic fields
are defined by ambient densities and velocity/magnetic fields
(ambient Alfv\'en Velocity) as well as Beltrami parameters
(helicities) -- such situation may be met in the poles of
magnetic WD's or for WDs with not detectable
surface magnetic fields.


\begin{figure}
\begin{center}
\includegraphics[scale=0.4,angle=0]{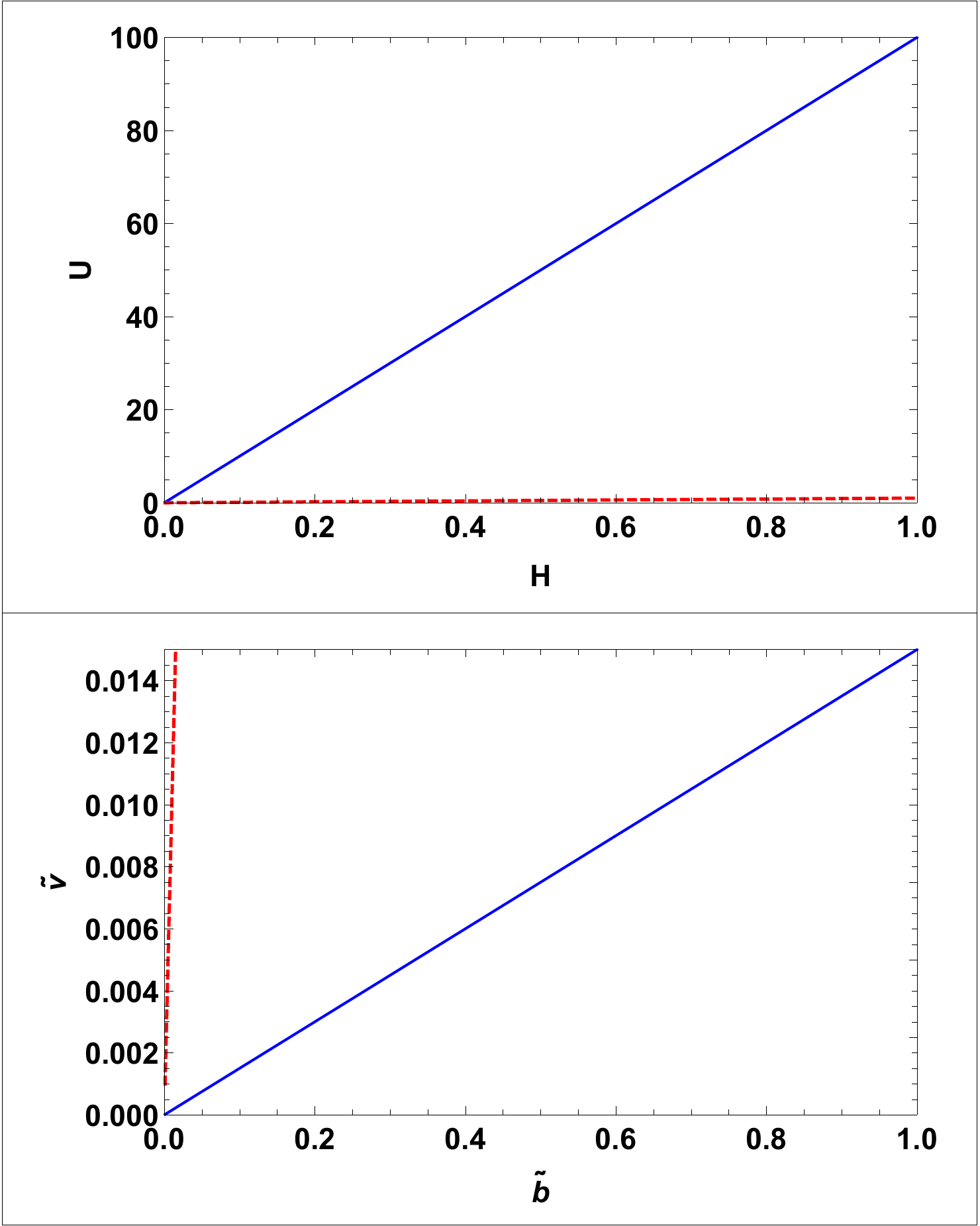}
\caption{Evolution of generated velocity and magnetic fluctuations (blue lines) for
Beltrami Parameneters $a\sim d = 0.01$: top panel shows the macro-scale
velocity field ${\boldsymbol{U}}$ versus macro-scale magnetic field
${\boldsymbol{H}}$ while the bottom panel shows the micro-scale
${\boldsymbol{\tilde{v}}}$ versus micro-scale ${\boldsymbol{\tilde{b}}}$. Starting from
sub-Alfv\'enic ambient micro-flow system arrives to strongly super-A;fv\'enic
macro-scale and sub-Alfv\'enic micro-scale fluctuation, locally. Reverse Dynamo
for macro--scale and Straight Dynamo for micro-scale. Dotted lines in red show the Alffv\'enic flow
for both scales, respectively.  }
\label{Fig.3.}
\end{center}
\end{figure}


\begin{figure}
\begin{center}
\includegraphics[scale=0.4,angle=0]{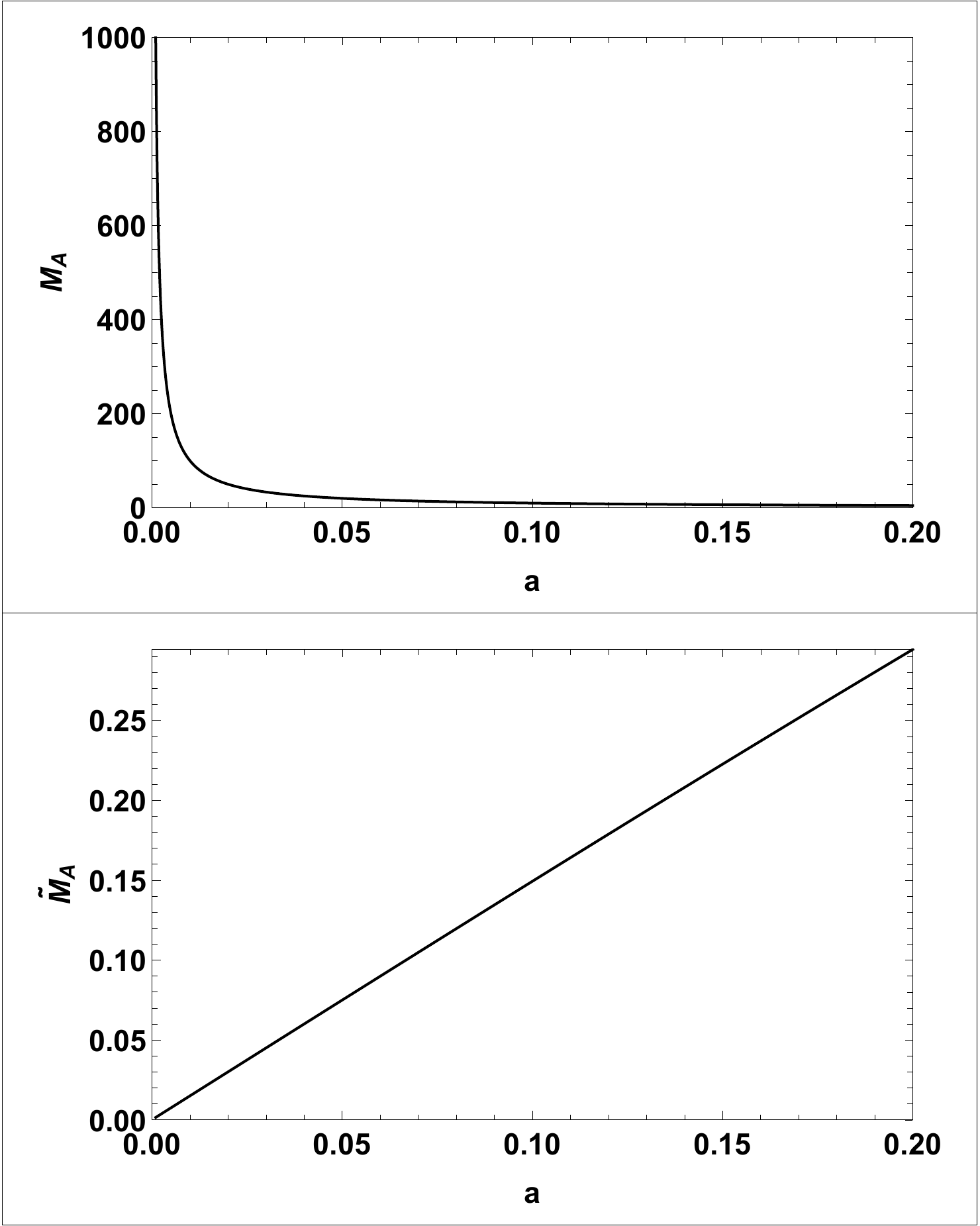}
\caption{Plot for Alfv\'en Mach Number versus $a < 1$ for macro-scale vector-fields
$M_A$ (top) and micro-scale vector-fields
${\tilde{M}}_A$ (bottom), respectively  for generated
velocity and magnetic fluctuations. Smaller the Beltrami scale $a$
bigger is the Macro-scale Alfv\'en Mach number that is $\gg 1$ for
all $a$-s in range while for micro-scale situation reverses --
Manifestation of Unified Reverse Dynamo/Dynamo mechanism. }
\label{Fig.4.}
\end{center}
\end{figure}

(2) Example of primarily magnetic ambient fields:
$a\sim d\ \ll 1 , (\lambda = a - a^{-1}\gg 1)$ implying
${\boldsymbol{v}}_0=(\lambda +a^{-1}){\boldsymbol{b}}_0\sim a\,\boldsymbol{b}_0
\ll \boldsymbol{b}_0$. These conditions maybe met in
WD's photospheres' certain convective areas where a
strongly sub-Alfve\'nic turbulent flow may exist.
This micro-scale magnetically dominant initial system
creates macroscale fields $\boldsymbol{U}\sim a^{-1}\boldsymbol{H}\gg
\boldsymbol{H}$ that are kinetically abundant -- Reverse Dynamo
scenario: in the given region of WD's photosphere
where the fluctuating/turbulent magnetic field is initially
dominant, the magneto-fluid coupling induces efficient/significant
acceleration, and part of the magnetic energy will be transferred
to steady plasma flows that are strongly super-Alfve\'nic; process is
accompanied by a weak macro-scale magnetic field generation.
In this regime generated micro-fields remain magnetically dominated,
hence defining the Unified Reverse Dynamo/Dynamo mechanism in its
full strength (see Fig.3 and Fig.4). It is interesting that this
mechanism works for any level of degeneracy and
may help us to predict the jet/outflow origin
in the atmospheres/outer layers of WD's.

\section{Model Equations for RD for Degenerate e-p Astrophysical Plasmas}

The dynamical phenomena in e-p plasmas develop differently
from their counterparts in the usual e-i system.
The annihilation, which takes place in the interaction of
electrons and positrons occurs at much longer characteristic
time scales compared with the time in which the collective
interaction between the charged particles takes place;
the details of such processes in super-dense e-p plasma
can be found in (Berezhiani et al. 2015b, and references therein).
Notice that even with equal effective masses
at equal temperatures, inertia change due to degeneracy can
cause asymmetry in e-p fluid \citep{Multi-B};
the degenerate e-p system is capable of creating
length scales larger than its classical counterpart
through the degeneracy enhanced inertia of the particles
\citep{SMB_multi}; this scale turns out to be degeneracy
dependent. Using this formalism, introducing the e-p plasma
bulk velocity as ${\bf V}  =\frac{1}{2}({\bf V}_+ + {\bf V}_-)$, assuming
$\gamma_- \sim \gamma_+ \sim 1; \ N_- = N_+ \equiv N$ leading to $G_- \equiv G_+
\equiv G_0(n) = const$ we derive the equations to describe the dynamics of
such system:
\[
\frac{\partial {\bf V}}{\partial t} = {\bf V} \times\nabla\times {\bf V} \
+ \ (\nabla\times {\bf b})\times {\bf b}
\]
\begin{equation}
\qquad \qquad +\ \alpha^2(\nabla \times {\bf b})\times \nabla \times \nabla \times {\bf b} \ ,
\label{WD13}
\end{equation}
\[
\frac{\partial{\bf b}}{\partial t} \ + \ \alpha^2\frac{\partial}{\partial t}
\nabla\times\nabla\times {\bf b} = \nabla \times ({\bf V}\times {\bf b})
\]
\begin{equation}
+ \ \alpha^2\nabla\times({\bf V}\times\nabla\times\nabla\times {\bf b}
\ + \ \nabla\times {\bf b}\times\nabla\times {\bf V}) \ ,
 \label{WD14}
\end{equation}
where all velocities are measured in terms of the corresponding
Alfv\'en velocity \ $V_A = \frac{B_0}{\sqrt{8\pi n_0 mG_0}}$ ,
all lengths  are normalized to some characteristic length $L$ (e.g. radius
of the compact star); $\alpha=\lambda_{\rm {eff}}/L $ with $\lambda_{\rm{eff}} $ being
an "effective" electron (positron) skin depth -- $\lambda_{\rm{eff}}=
\frac{1}{\sqrt{2}}\frac{c}{\omega_p}=c\sqrt{\frac{mG_0}{8\pi n_0e^2}}$.
In such degenerate e-p plasma
the equilibrium state constiututes the Triple-Beltrami structures
with 3 different scales. Then, following the standard procedure,
introducing for velocity and magnetic fields similar to (\ref{WD3})
representations, we use equilibrium equations from \citep{SMB_multi}:
\[
2G_0^2\nabla\times\nabla\times\nabla\times{\bf b}_0
-2G_0(a_+ - a_-)\,\nabla\times\nabla\times {\bf b}_0
\]
\begin{equation}
+ 2G_0(G_0-a_+a_-)\,\nabla\times {\bf b}_0
-(a_+-a_-)\,{\bf b}_0 = 0 \ ,
\label{WD15}
\end{equation}
\begin{equation}
{\bf V}_0 = (a_+ + a_-)^{-1}(2G_0\nabla\times\nabla\times {\bf b}_0)
\label{WD16}
\end{equation}
\[
\qquad - (a_+ + a_-)^{-1}\left[(a_+-a_-)\nabla\times{\bf b}_0
+ 2{\bf b}_0\right] \ ,
\]
where $a_{\pm}$ are dimensionless constants related to
two invariants:  \ $h_{\pm} = \int{({\bf A}\pm G_0{\bf V}_{\pm})\cdot
({\bf b}\pm G_0\nabla \times {\bf V}_{\pm})\,d^3x}$
-- the generalized helicities of the system. Here the
generalized vorticities for both fluids have now both
magnetic and kinetic parts due to the degeneracy effects.
We need to choose these constants so that the scales
[solutions of the equation \ $2G_0^2\,\mu^3 - 2G_0(a_+ - a_-)\,
\mu^2 + 2G_0(G_0-a_+a_-)\,\mu -(a_+-a_-)= 0$]
\ are vastly separated. For the analysis we choose the simplest
case when $a_+ \sim a_- \simeq a$, then the three real roots are: \
$0; \ \pm \sqrt{a^2 - G_0}/G_0$ \ with the condition $a^2 > G_0$.
Following the similar to Section 2 procedure we assume that
the basic reservoir is fully micro-scale leading to the
velocity and magnetic fields being linearly related as
${\bf V}_0 = \frac{a}{G_0}{\bf b}_0$ when choosing the inverse
micro-scale to be $\lambda = \sqrt{a^2 - G_0}/G_0$.
Note, that when $a > G_0$ the ambient micro-scale fields are
primarily kinetic, while at $\sqrt{G_0}<a<G_0$
they are primarily magnetic.

The straightforward algebra leads to the following equations
for micro-scale and macro-scale fluctuations [${\bf Q}$ and
${\bf S}$ are the functions containing ${\b U}$ and ${\b H}$]:
\begin{equation}
\frac{\partial{\tilde{\bf b}}}{\partial t}=({\bf Q}\cdot \nabla) {\bf b}_0 \ ,
\qquad \frac{\partial{\tilde{\bf V}}}{\partial t}
= ({\bf S}\cdot \nabla)\,{\bf b}_0 \ ,
\label{WD17}
\end{equation}
and
\begin{equation}
\ddot{\bf H} = r \nabla \times {\bf H} + m \nabla\times\nabla\times {\bf H}
+ \nu \nabla \times \nabla \times {\bf U} \ ,
\label{WD18}
\end{equation}
\begin{equation}
\ddot{\bf U} = s \nabla \times {\bf U} + q\nabla\times {\bf H}
+ l \nabla \times \nabla \times {\bf U} + p \nabla\times\nabla\times {\bf H}
\label{WD19}
\end{equation}
where:
\[
r \equiv \frac{\lambda b_0^2}{3}\left[\left(\frac{a}{G_0}\right)^2-1-\lambda^2\right],
\quad l \equiv - \alpha \,\frac{\lambda^2 b_0 ^2}{6}(2+3\lambda ^2) ,
\]
\[
\nu \equiv \alpha \,\frac{\lambda^2 b_0^2}{6}\left(\frac{a}{G_0}\right)\,(3+2\lambda^2) ,
\ \ q \equiv -\left(\frac{a}{G_0}\right)\frac{\lambda^3 b_0 ^2}{3} ,
\]
\[
p \equiv \alpha\,\frac{\lambda^2 b_0^2}{3}\left(\frac{a}{G_0}\right)(1+\lambda^2) ,
\ \ \nu = p + \alpha\,\frac{\lambda^2 b_0 ^2}{6}\left(\frac{a}{G_0}\right) ,
\]
\[
s \equiv \frac{\lambda b_0 ^2}{6}\left[1-\left(\frac{a}{G_0}\right)^2+3\lambda^3+2\lambda^4\right] ,
\]
\begin{equation}
m \equiv \alpha\frac{\lambda^2 b_0^2}{6}\left[1-4\left(\frac{a}{G_0}\right)^2\right] \ .
\label{WD20}
\end{equation}

Performing a Fourier analysis we obtain the following
dispersion relation:
\begin{equation}
\qquad \qquad D^2(\omega , k)= C^2(\omega , k)\,k^2 , \quad {\rm with}
\label{WD21}
\end{equation}
\[
D(\omega ,k) = \Big((\omega^2+lk^2)(\omega^2+mk^2)+rsk^2-p\nu k^4\Big) ,
\]
\[
C(\omega ,k) = \Big((\omega^2+lk ^2)r+(\omega^2+mk^2)s-q\nu k^2\Big) ,
\]
and, finally, the relation for macro-scale fluctuations:
\begin{equation}
\qquad \qquad {\bf U} = \frac{[r\,C- (mk^2 - \omega^2)\,D]}{\nu k^2\,D}\,{\bf H} \ .
\label{WD22}
\end{equation}

We observe that to leading order, unlike the degenerate e-i case,
the evolution of ${\bf H}$ does require knowledge of ${\bf U}$
and vice versa; a choice of Beltrami parameter $a$
(that now reflects the helicities of degenerate e-p system) as well as the
density (through effective mass $G_0$), fixes the relative amounts
in ambient fields' microscopic energy and, consequently,
in the generated macroscopic fields that grow proportionately to each other.
Below we show how the Unified RD/Dynamo mechanism
affects the evolution picture of astrophysical objects
with degenerate e--p plasmas.

\subsection{Unified Dynamo/Reverse Dynamo mechanism for
Degenerate e-p astrophysical plasmas}

As discussed above, super-dense e–-p astrophysical plasma density
is argued to be in the range $n = (10^{30} –- 10^{37})\,cm^{-3}$.
E.g. the effective mass $G_0 \sim 25$ for the density $\sim 10^{34}\,cm^{-3}$.
Then, for such objects we can examine the two extreme cases for
Beltrami parameter: (1) $a \sim 100 \gg G_0$ and (2)
$ \sqrt{G_0} < a\sim 10 < G_0$. Since $\alpha $ - Hall term
contribution - is very small (in the simulations we are using
$\alpha \sim 10^{-6}$ ) the coefficients $l, \ p, \ q, \ m, \nu $
are normally small; coefficients $r$ and $s$ are free of
$\alpha $, so they may become the determining ones in the
dispersion as well as in the ratio for generated fluctuations.
Also, since inverse micro-scale $\lambda = \frac{a}{G}\sqrt{1
- \frac{G_0}{a^2}}$, we have $r < 0$ in our analysis
(defining the growth rate of generated macro-fields via
dispersion relation).


\begin{figure}
\begin{center}
\includegraphics[scale=0.33,angle=0]{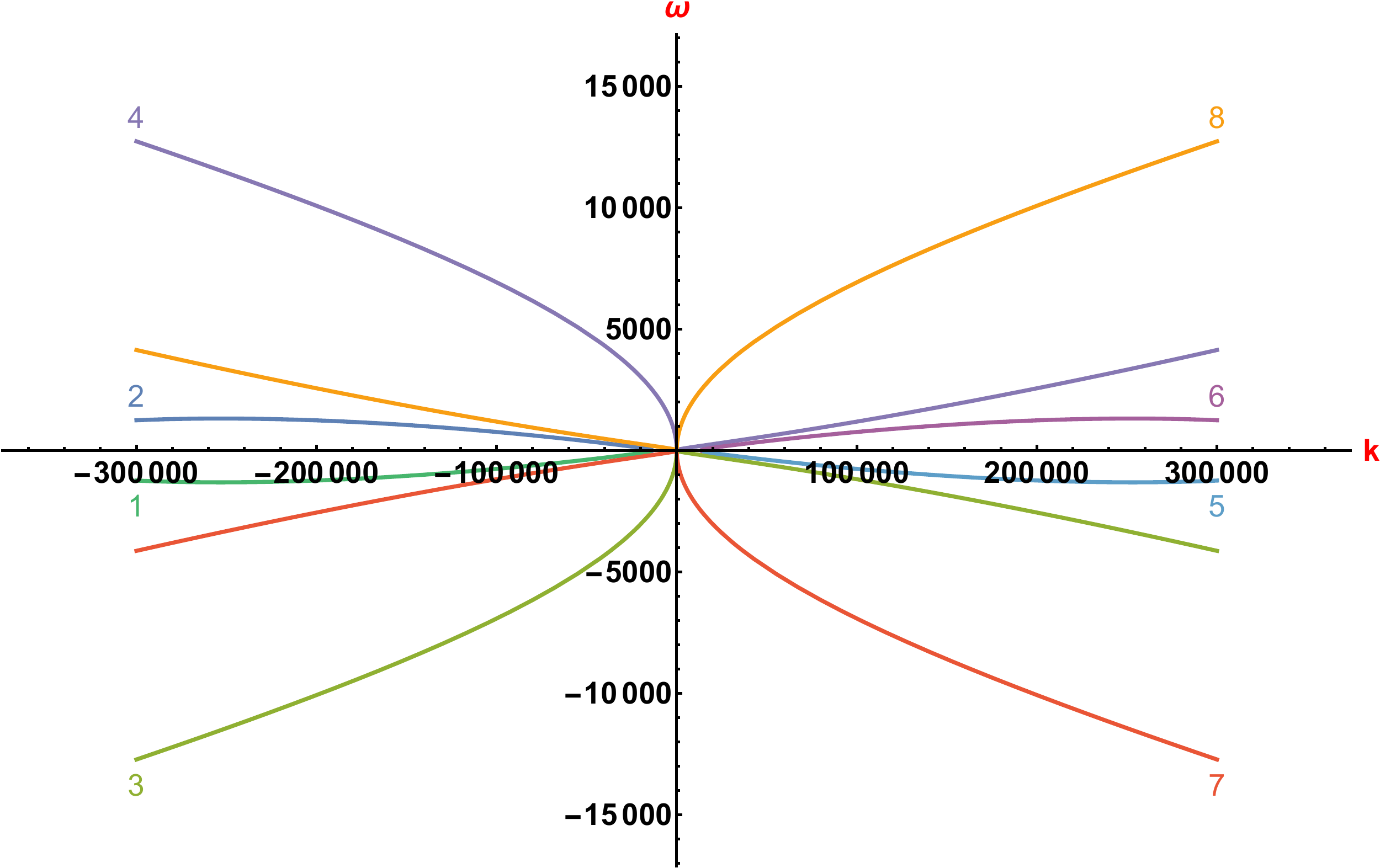}
\caption{Solution of dispersion relation (\ref{WD21})
for $a=100, G_0=25$ ; 8 different real roots
are displayed by different color. }
\label{Fig.5.}
\end{center}
\end{figure}


\begin{figure}
\begin{center}
\includegraphics[scale=0.4,angle=0]{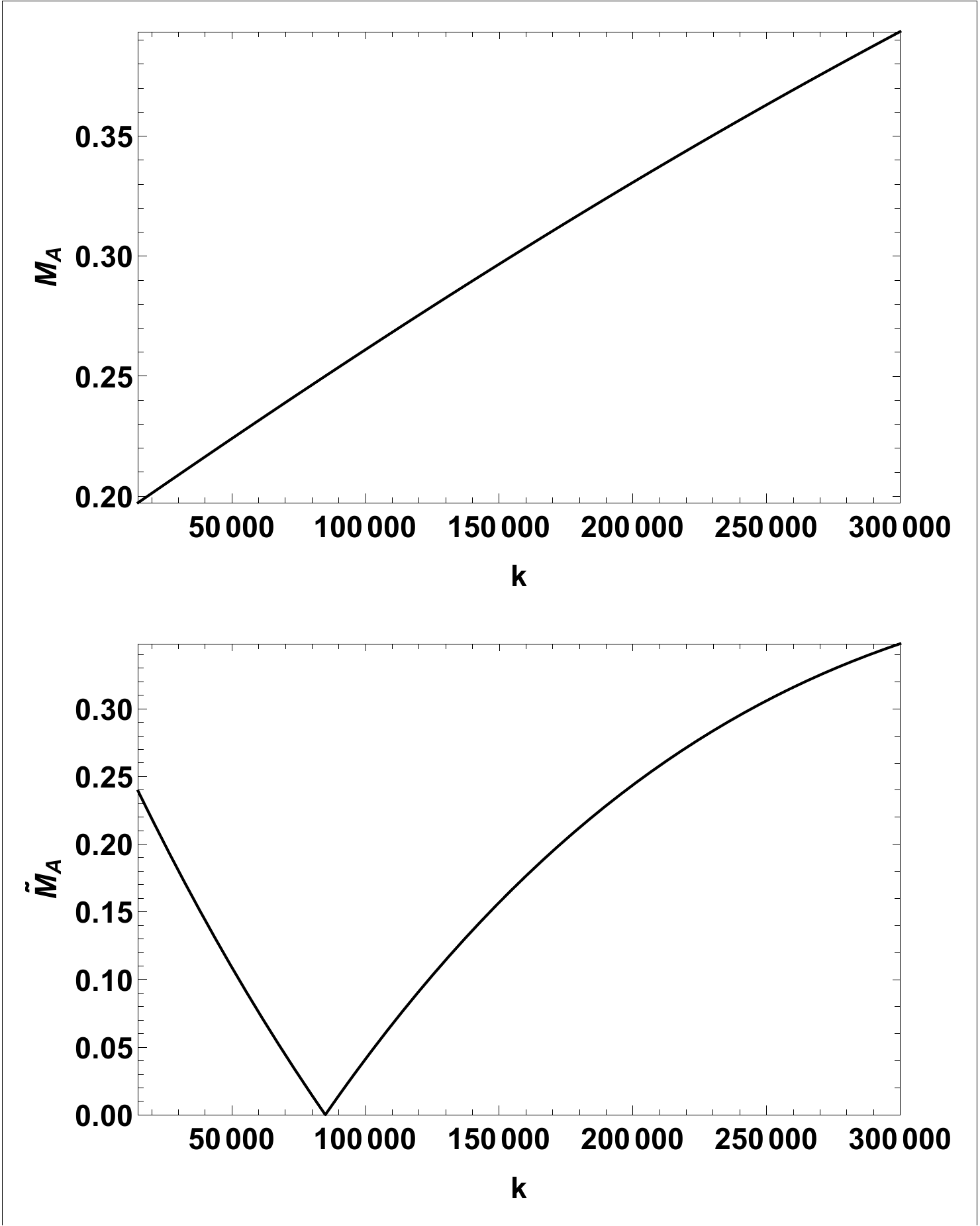}
\caption{Plot for Alfv\'en Mach Numbers versus $k$ for:
macro-scale vector-fields $M_A$ (top) and micro-scale vector-fields
${\tilde{M}_A}$ (bottom), respectively  for generated
velocity and magnetic fluctuations for the root 6 of Fig.5;
$a=100 > G_0=25$. Bigger the $k$ smaller is $M_A$
but still $\ll 1$ ; both scale
fluctuations are Sub-Alfv\'enic; $\alpha = 10^{-6}$.}
\label{Fig.6.}
\end{center}
\end{figure}


\begin{figure}
\begin{center}
\includegraphics[scale=0.4,angle=0]{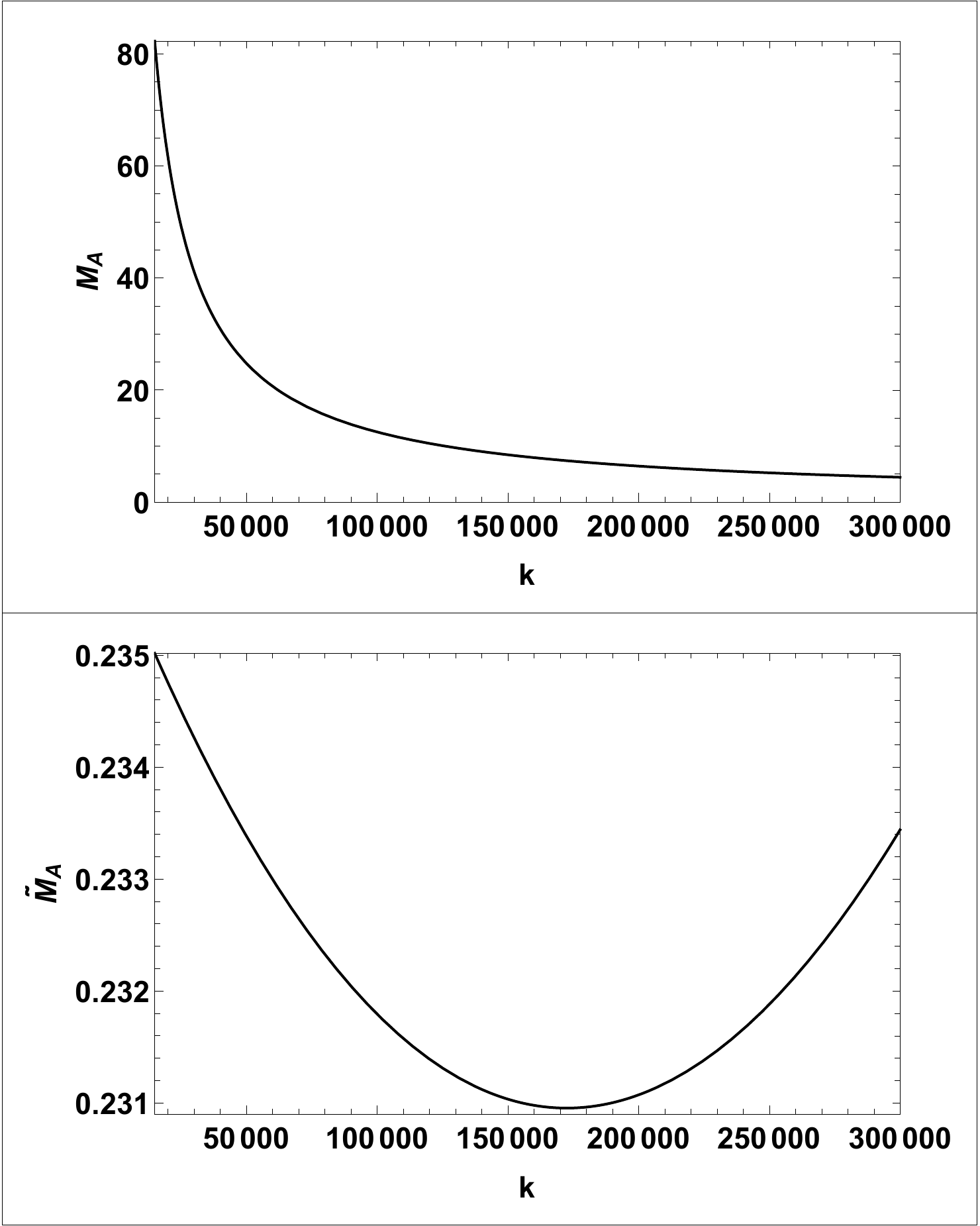}
\caption{Plot for Alfv\'en Mach Numbers versus $k$ for:
macro-scale vector-fields $M_A$ (top) and micro-scale vector-fields
${\tilde{M}_A}$ (bottom), respectively  for generated
velocity and magnetic fluctuations for the root 8 of Fig.5;
$a=100 > G_0=25$. Smaller the $k$ bigger is $M_A$ that is $\gg 1$
-- manifestation of Unified Reverse Dynamo/Dynamo mechanism; $\alpha = 10^{-6}$.}
\label{Fig.7.}
\end{center}
\end{figure}

(1). For $a \sim 100 \gg G_0$  and
${\bf V}_0 = \frac{a}{G_0}{\bf b}_0 \gg {\bf b}_0$ --
ambient flow is primarily kinetic. Dispersion relation
is solved numerically, corresponding 8 real roots are
displayed in Fig.5. In Fig.6 Alfv\'en Mach numbers
for both the generated macro-- and micro-fields is
displayed for root 6. We observe, that both--scale
generated velocity fields are Sub-Alfv\'enic.
This is straight Dynamo scenario predicting the
strong magnetic field generation simultaneously to
weak flows/outflows. This may explain the existence
of strong magnetic fields in the vicinity of massive stars.
In Fig.7 Alfv\'en Mach numbers are plotted for root 8, for
which, interestingly, there is a Unified RD/Dynamo
process when Macro-Scale flow is Super-Alfv\'enic
and short-scale fluctuations follow Dynamo.


\begin{figure}
\begin{center}
\includegraphics[scale=0.22,angle=0]{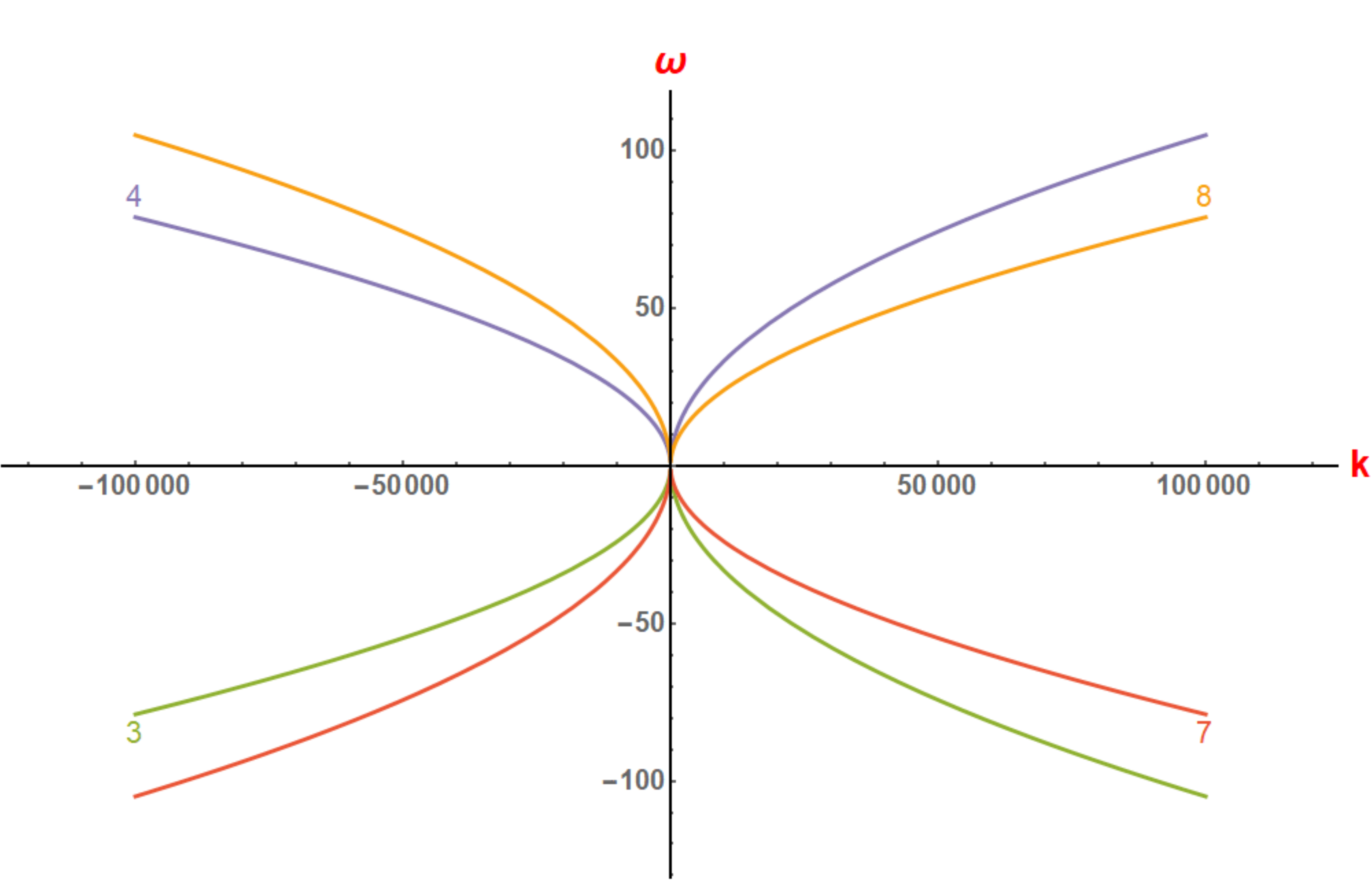}
\caption{Solution of dispersion relation (\ref{WD21})
for $a =10, G_0=25$ ; 4 different real roots
are displayed by different color. }
\label{Fig.8.}
\end{center}
\end{figure}


\begin{figure}
\begin{center}
\includegraphics[scale=0.4,angle=0]{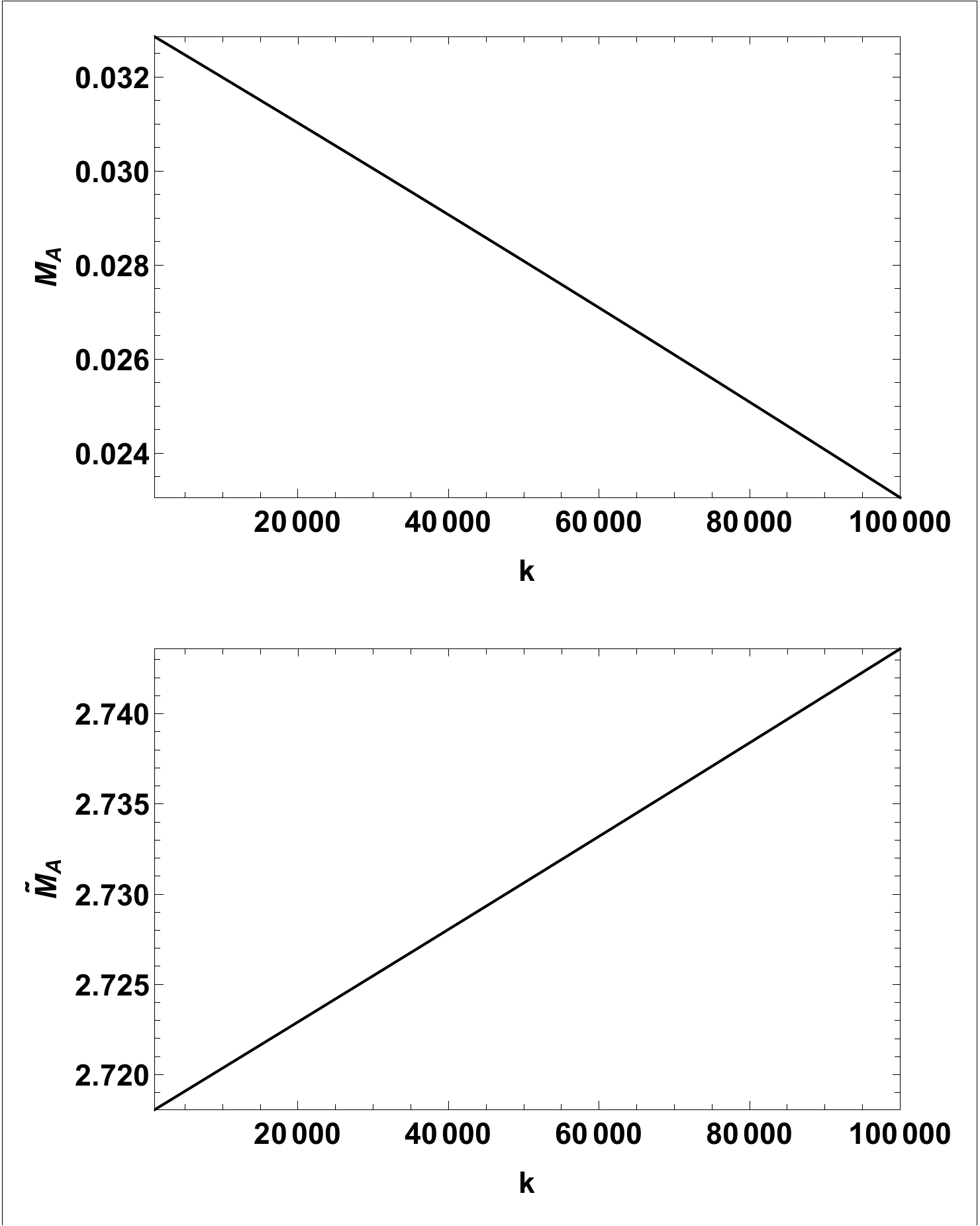}
\caption{Plot for Alfv\'en Mach Numbers versus $k$ for:
macro-scale vector-fields $M_A$ (top) and micro-scale vector-fields
${\tilde{M}_A}$ (bottom), respectively  for generated
velocity and magnetic fluctuations for the root 4 of Fig.8;
$a=10 > \sqrt{G_0}=5$. Bigger the $k$ smaller is $M_A$
but still $\ll 1$ ; micro scale fluctuations are Super-Alfv\'enic.
$\alpha = 10^{-6}$ .}
\label{Fig.9.}
\end{center}
\end{figure}


\begin{figure}
\begin{center}
\includegraphics[scale=0.4,angle=0]{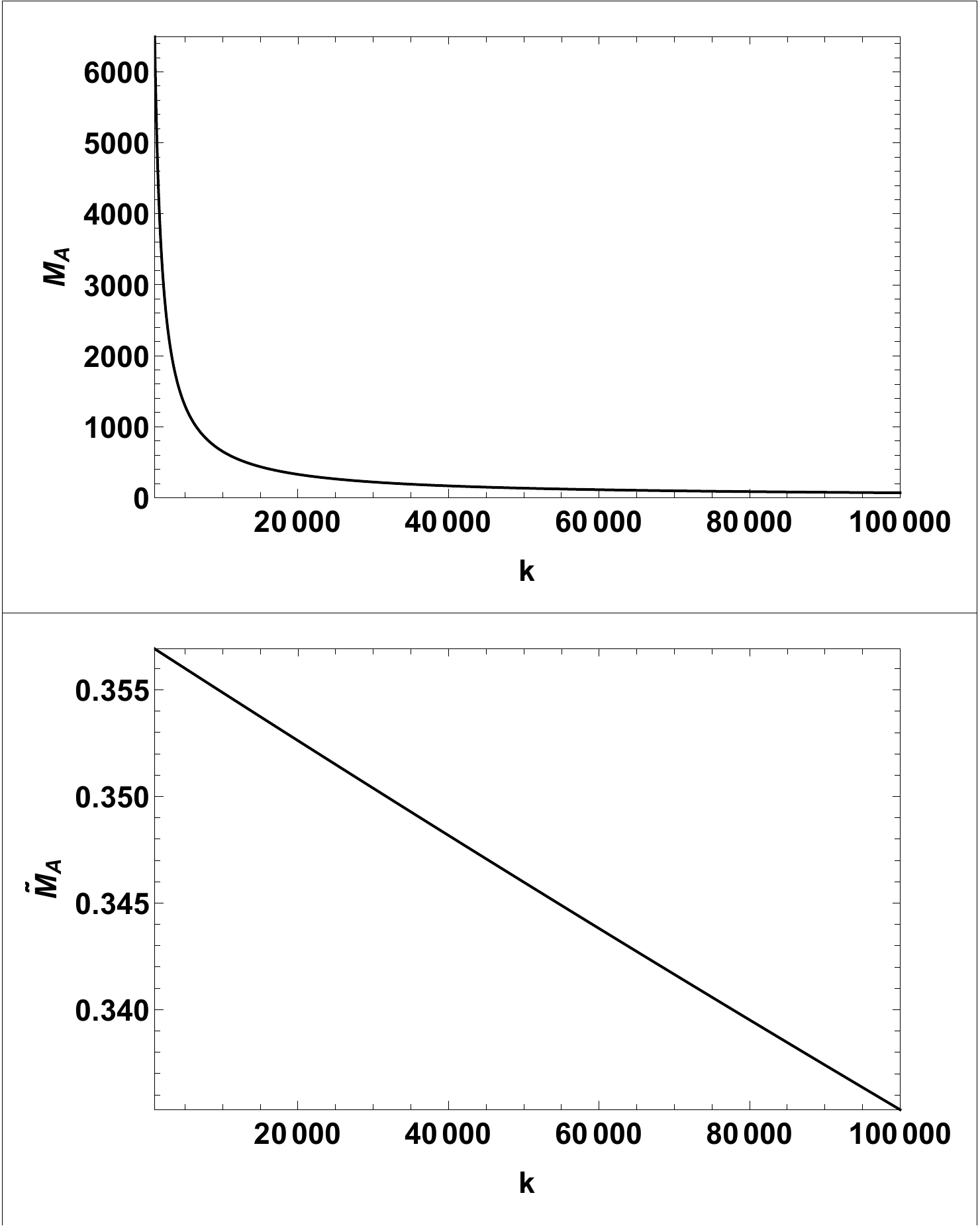}
\caption{Plot for Alfv\'en Mach Numbers versus $k$ for:
macro-scale vector-fields $M_A$ (top) and micro-scale vector-fields
${\tilde{M}_A}$ (bottom), respectively  for generated
velocity and magnetic fluctuations for the root 8 of Fig.8;
$a=10 > \sqrt{G_0}=5$. Smaller the $k$ bigger is the $M_A$ that
is $\gg 1$  while for micro-scale situation for Mach number reverses --
Manifestation of Unified RD/Dynamo. $\alpha = 10^{-6}$.}
\label{Fig.10.}
\end{center}
\end{figure}

(2) For $a \sim 10 \gg \sqrt{G}$,  and
${\bf V}_0 = \frac{a}{G}{\bf b}_0 \ll {\bf b}_0$ --
ambient flow is primarily magnetic. Dispersion
relation is solved numerically and corresponding 8
real roots are displayed in Fig.8. In Fig.9 Alfv\'en
Mach numbers for both the generated macro-- and
micro-fields are displayed for root 4. We observe,
that the generated macro-scale flows are sub-Alfv\'enic
while micro-scale flows are super-Alfve\'nic. This is
the illustration of Unified Dynamo/RD scenario predicting
strong magnetic field generation simultaneously
to flows/outflows that are sub-Alfv\'enic. This, again, may explain
the existence of strong magnetic fields in the vicinity of
massive stars. Such scenario was absent in
degenerate e-i case (see results of Section 2).
The RD at short-scales for general settings
was discussed in \citep{brand}. In Fig.10 Alfv\'en
Mach numbers are plotted for root 8 for which
we again observe a Unified RD/Dynamo process
but now the generated macro-scale super-Alfv\'enic flow is very fast;
short-scale fluctuations follow Dynamo process. According
to \citep{beskin} $M_A\geq 10^3$ is observed in
a variety of astrophysical outflows. It is remarkable
that our result shows similar rate extremely fast flows for
small $k$-s.


\begin{figure}
\begin{center}
\includegraphics[scale=0.45,angle=0]{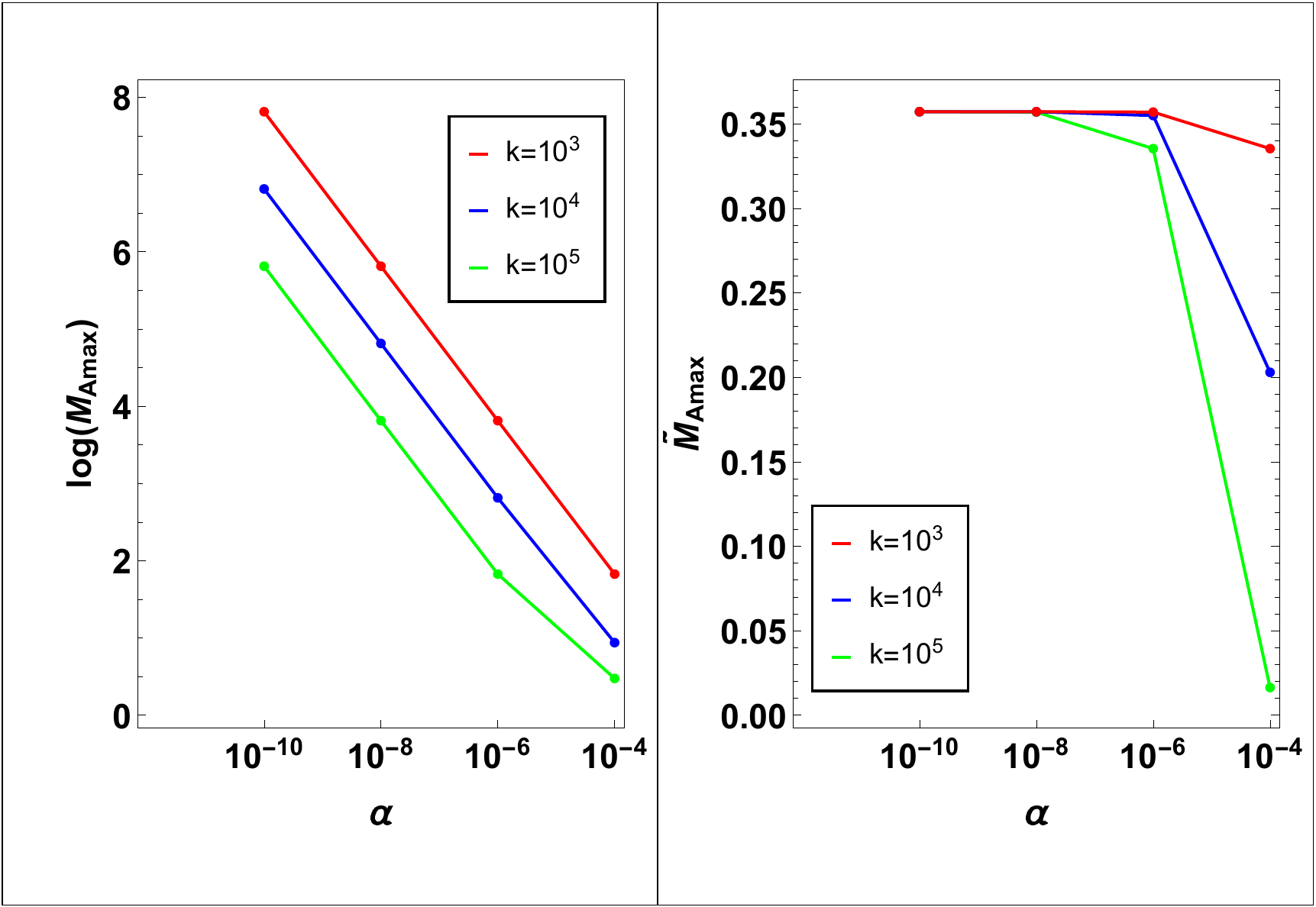}
\caption{Maximal values of Alfv\'en Mach Numbers versus $\alpha$
for: macro-scale vector-fields $M_{Amax}$ (left) and micro-scale
vector-fields ${\tilde{M}_{Amax}}$ (right), respectively for generated
velocity and magnetic fluctuations (for the root 8 of Fig.8);
$a=10 > \sqrt{G_0}=5$. Smaller the $\alpha $ bigger is the $M_{Amax}$ that
can reach values $\gg 10^3$  at small $k$ (see Eq. (\ref{WD22})).
Red, blue, green colors correspond to:
$k=10^3 , \ 10^4 , \ 10^5$, respectively. Smaller the $k$ bigger is $M_{Amax}$
for the same $\alpha$.}
\label{Fig.11.}
\end{center}
\end{figure}

We have examined the dependence of final results
on the dimensionless parameter $\alpha \ll 1$. Results
presented in Fig.11 show that maximal macro-scale Alfv\'en Mach
number $M_{Amax}$ increases when decreasing $\alpha $ and can
reach the values $\sim 10^8$ for very small $\alpha \lesssim 10^{-10} $
at $k \lesssim 1000$.  This result proves that generated outflow
has dispersion with ${\b U}({\b r},t)$ macro-scale velocity field
varying in space and time. Our detailed analysis shows that for
smaller densities (hence, for weaker degeneracy level $G_0$) picture changes
very slightly, maximal values of macro-scale outflow reduce
a little, while maximal values of micro-scale outflow increase
more significantly. Main result is the same -- there is a
fast flow/outflow formation in degenerate e-p plasma of astrophysical
objects due to Unified Reverse Dynamo/Dynamo mechanism. We understand,
that when heating as well as well wave processes are included
in the dynamical process of acceleration final values may change
according to additional channels of energy transformations.

Thus, analysis showed that for strongly degenerate e-p astrophysical
plasma the Unified RD/Dynamo scenario is guaranteed
for one of the roots of dispersion relation; depending on
the range of Beltrami scales (helicities) of e-p fluids
as well as density ($G_0$ - degeneracy level) the ratio
between macro-scale velocity and magnetic fields may become different;
similar conclusions can be drawn for micro-scale fields.
At the end, there can be any mixture of macro-- and micro-- fields
over the time. There will be conditions favorable for macro-scale
super-Alfv\'enic flows to be generated together with the
macro-scale magnetic fields and from these macro-scale
solutions one can extract the density (degeneracy level)
(see \citep{lingam} for details).

\section{Conclusions}

From an analysis of the degenerate two-fluid system, we
have extracted the Unified Reverse Dynamo/Dynamo mechanism
-— the amplification/generation of fast macro-scale plasma
flows in astrophysical systems with initial turbulent
(micro--scale) magnetic/velocity fields. This process is
simultaneous with and complementary to the micro-scale
unified Dynamo/RD dynamics. It is found (both analytically and numerically)
that like in the classical case the generation of macro--scale
flows is an essential consequence of the magneto-fluid coupling.
The generation of macro--scale fast flows and magnetic fields are simultaneous:
the greater the macro--scale magnetic field (generated locally)
the greater becomes the macro--scale velocity field (generated
locally). Principle results of our investigation are following:
\begin{itemize}
\item {The flow/outflow acceleration due to the Unified RD/Dynamo
is directly proportional to the initial turbulent magnetic
energy in degenerate e-i astrophysical plasma, while in degenerate e-p
plasma such flows are fed by either initial turbulent state:
kinetically or magnetically dominated ambient system;}
in the latter case generated/accelarated outflows
are very strong.

\item{Depending on ambient density ($G_0(n)$)
of degenerate two-fluid system the scenarios become different
for different Beltrami parameter ($a$) but there always
exists such a real $\omega $ (solution of dispersion relation)
for which the generation of strong macro-scale fast locally
Super-Alfv\'enic flow/outflow is guaranteed. }

\item{Formation process
is very sensitive to both the degeneracy state of the system and
the magneto-fluid coupling. For the same parameters ($G_0$ and $a$),
generated flow/outflow will have a dispersion with
Velocity field distribution in (${\bf r}, t$) --
observations show that large-scale astrophysical flows are
very complex with a characteristic evolution in time and space.}

\item{It is interesting, that even the accelerated flows are sub-Alfv\'enic
in some regimes for ambient system parameters, the flows, along with
the dominant magnetic fields, will continue to amplify as long as
there is an ambient turbulent energy to drive them -- both fields
grow at the rate defined by the dispersion relation (\ref{WD21})
and determining parameters are $\alpha$ and $r$ (defined by (\ref{WD20})).}

\item{In case of degenerate e-i plasma when the microscopic magnetic field
is initially dominant, a major part of its energy transforms to
super-fast super-Alfv\'enic macro--scale outflow energy
due to magneto-fluid coupling; a weak macro--scale
magnetic field is generated along with it. Specifically
important finding is that in degenerate e-p plasma the
whole set of cases may exist because of the different
channels related to different real roots of dispersion relation,
and for realistic physical parameters the
resulting accelerated / generated locally super-Alfv\'enic flows
are extremely fast with Alfv\'en Mach
number $> 10^3$ as observed in a variety of astrophysical outflows.}
\end{itemize}

Thus, the unified Reverse Dynamo/Dynamo mechanism, providing an unfailing source
for macro-scale fast flows/outflows becomes very significant together with
other additional mechanisms (e.g. energy transformations due to catastrophe or waves)
for understanding the existence of
fast macro-scale flows/outflows in astrophysical objects with
degenerate components -- there is an intrinsic tendency of
flow/magnetic field amplification due to magneto-fluid coupling
in such system.

\section{Acknowledgements}

Authors express their thanks to Dr. Alexander Barnaveli for
the valuable discussions. This work was supported by
Shota Rustaveli Georgian National Foundation Grant Project
No. FR17-391.


%



\begin{thebibliography}{}





\bibitem[Aksenov et al 2010]{aksenov} Aksenov, A.G., Ruffini, R. and Vereshchagin, G.V.
Phys. Rev. E {\bf 81} 046401 (2010).

\bibitem[Arshilava et al. 2019]{yso} Arshilava, E. Gogilashvilia, M., Loladzea, V.,
Jokhadzea, I., Modrekiladzea, B., Shatashvilia, N.L. Tevzadze, A.G.
J. High Energy Astrophysics {\bf 23}, 6 (2019).

\bibitem[Barnaveli \& Shatashvili 2017]{BS-flow} Barnaveli, A.A., Shatashvili, N.L.
Astrophys Space Sci. 362:164 (2017).

\bibitem[Begelman et al 1984]{Begelman} Begelman, M.C., Blandford, R.D., and
Rees, M.D. \ Rev. Mod. Phys. {\bf 56} 255 (1984).

\bibitem[Beloborodov \& Thompson 2007]{Beloborodov} Beloborodov, A.M.  and
Thompson, C. \apj \textbf{657}, 967 (2007).

\bibitem[Berezhiani et al 2015a]{BSM_deg} Berezhiani, V.I., Shatashvili, N.L. and Mahajan, S.M.
Phys. Plasmas {\bf 22}, 022902 (2015a).

\bibitem[Berezhiani et al 2015b]{degenerate} Berezhiani, V.I., Shatashvili, N.L. and
Tsintsadze, N.l. Physica Scripta {\bf 90(6)}, 068005 (2015b).

\bibitem[Beskin 2010]{beskin} Beskin V. S.,Phys.-Usp., {\bf 53}, 1199 ( 2010).

\bibitem[Bodo et al 2015]{bodo} Bodo, G.; Cattaneo, F.; Mignone, A.; Ponzo, F.; Rossi, P. ApJ, 808(2),
141, (2015)

\bibitem[Branderburg \& Rempel 2019]{brand} Brandenburg, A., Rempel, M. \apj, {\bf 879}, 57 (2019).

\bibitem[Cercignani \& Kremer 2002]{Russo-2} Cercignani, C. and Kremer, G.M. \ 2002
{\it The relativistic Boltzmann equation: theory and applications}
Birkh\"{a}user, Basel; chapter 3.

\bibitem[Chandrasekhar 1931]{Chandra1} Chandrasekhar, S. \apj {\bf 74}, 81 (1931);
Mon. Not. R. Astron. Soc. 95, 207 (1935).

\bibitem[Chandrasekhar 1939]{Chandra2} Chandrasekhar, S. \textit{An Introduction to the Study of
Stellar structures}, Chicago (Dover Publications, 1939).

\bibitem[Cumming 2002]{cumming} Cumming A., MNRAS, {\bf 333}, 589 (2002).

\bibitem[Dunne 2006]{Dunne-1} Dunne, M. \  A high-power laser fusion facility for
Europe \ Nature Phys. {\bf 2} 2 (2006).

\bibitem[Ferrario et al 2015]{ferrario} Ferrario, L., de Martino, D.,
\& Gaensicke, B. T. SSRv, 1 (2015).

\bibitem[Gedalin et al 1998]{Misha} Gedalin, M., Melrose, D.B, and Gruman, E. \
Phys. Rev. E {\bf 57} 3399 (1998).

\bibitem[Han et al 2012]{ruffini} Han, W.B., Ruffini, R. and Xue, S.S. \
\prd {\bf 86} 084004 (2012).

\bibitem[Hollands et al 2015]{hollands} Hollands, M., Gaensicke, B.,
\& Koester, D. MNRAS, {\bf 450}, 68 (2015).

\bibitem[Kawka et al 2007]{Kawka} Kawka, A., Vennes, S., Schmidt, G. D.,
Wickramasinghe, D. T., \& Koch, R. \apj, {\bf 654}, 499 (2007).

\bibitem[Iqbal et al 2008]{iqbal-3} Iqbal, N., Berezhiani, V.I. and Yoshida, Z.
Phys. Plasmas {\bf 15}, 032905 (2008).

\bibitem[Kawka \& Vennes 2014]{DAZ} Kawka, A. and Vennes, S. MNRAS {\bf 439}, L90 (2014).

\bibitem[Kepler et al 2013]{kepler} Kepler, S. O., Pelisoli, I.,
Jordan, S., Kleinman, S.J., Koester, D., K\"ulebi, D.B., Pecanha, B.V., Castanheira, B.G.,
Nitta, A., Costa, J.E.S., Winget, D.E., Kanaan, A. and Fraga, L. MNRAS {\bf 429}, 2934
(2013).

\bibitem[Kissin \& Thommpson 2015]{kissing} Kissin, Y.,
\& Thompson, C. \apj, {\bf 809}, 108 (2015).

\bibitem[Koester \& Chanmugam 1990]{White} Koester, D. and
Chanmugam,G. Rep. Prog. Phys. \textbf{53}, 837 (1990).

\bibitem[Krishan \& Yoshida 2006]{vinod} Krishan, V. and Yoshida, Z.
Phys. Plasmas, {\bf 13(9)}, 092303 (2006).

\bibitem[K\"ulebi et al 2009]{kulebi} K\"ulebi, B., Jordan, S.,
Euchner, F., Gänsicke, B. T., \& Hirsch, H. A\&A, {\bf 506}, 1341 (2009).

\bibitem[Liebert et al 2003]{hmfwd} Liebert, J., Bergeron, P.,
\& Holberg, J. B. AJ, {\bf 125}, 348 (2003).

\bibitem[Lingam \& Mahajan 2015]{lingam} Lingam, M., Mahajan, S.M.
\mnras 449, L36–L40 (2015).

\bibitem[Mahajan \& Yoshida 1998]{MY} Mahajan, S.M.  and Yoshida, Z.
\prl {\bf 81}, 4863 (1998).

\bibitem[Mahajan et al 2001]{mmns-1} Mahajan, S.M., Miklaszewski, R.,
Nikol'skaya, K.I. and Shatashvili, N.L. Phys. Plasmas {\bf 8},
1340 (2001).

\bibitem[Mahajan et al 2002]{mnsy}  Mahajan, S.M., Nikol'skaya, K. I.,
Shatashvili,  N.L.  and Yoshida, Z. \apj {\bf 576}, L161 (2002).

\bibitem[Mahajan et al 2005;2006]{msms}  Mahajan, S.M., Shatashvili, N.L.,
Mikeladze, S.V. and Sigua, K.I. \apj {\bf 634}, 419 (2005); Phys.
Plasmas {\bf 13}, 062902 (2006).

\bibitem[Mahajan \& Lingam 2015]{Multi-B} Mahajan, S.M. and Lingam, M. Phys. Plasmas,
{\bf 22(9)}, 092123 (2015).

\bibitem[Mahajan \& Yoshida 2010]{RelVorticity} Mahajan, S.M. and
Yoshida, Z. \prl, {\bf 105}, 095005 (2010).

\bibitem[Michel 1991]{Sturok-3} Michel, F.C.  \textit{Theory of Neutron Star Magnetospheres},
University of Chicago Press, Chicago, (1991).

\bibitem[Michel 1982]{Compact-2} Michel, F.C. Rev. Mod. Phys. {\bf 54}, 1 (1982).

\bibitem[Minini et al. 2003]{minini} Mininni, P.D., Gomez, D.O.,
\& Mahajan, S.M. \apj {\bf 567}, L81 (2002).

\bibitem[Mourou et al 2006]{Dunne-2} Mourou, G.A., Tajima, T. and
Bulanov, S. V. \ {\it Rev. Mod. Phys.} {\bf 78} 309 (2006).

\bibitem[Ohsaki et al 2001,2002]{osym} Ohsaki, S., Shatashvili, N.L.,
Yoshida, Z. and Mahajan, S.M. \apj {\bf 559}, L61 (2001); \
\apj {\bf 570}, 395 (2002).

\bibitem[Pino et al 2010]{pino} Pino, J., Li, H. and Mahajan, S.M.
Phys. Plasmas {\bf 17}, 112112 (2010).

\bibitem[Ruderman \& Sutherland 1973]{ruderman} Ruderman, M.A.,
\& Sutherland, P.G. NPhS, {\b 246}, 93 (1973).

\bibitem[Schmidt et al 2003]{schmidt} Schmidt, G. D., Harris, H. C., Liebert, J.,
et al. \apj , {\bf 595}, 1101 (2003).

\bibitem[Shapiro \& Teukolsky 1973]{Compact-1} Shapiro, L. and
Teukolsky, S.A. {\it Black Holes, White Dwarfs and Neutron Stars:
The Physics of Compact Objects}, (John Wiley and Sons, New York,
1973).

\bibitem[Shatashvili et al 2016]{SMB_multi} Shatashvili, N.L., Mahajan, S.M.
and Berezhiani, V.I. AAS, {\bf 361}, 70 (2016).

\bibitem[Shatashvili \& Yoshida 2011]{SY-DJ} Shatashvili, N.L.  and Yoshida, Z.
AIP Conf. Proc. {\bf 1392}, 73 (2011).

\bibitem[Shiraishi et al 2009]{Shiraishi} Shiraishi, J, Yoshida, Z. and Furukawa, M.
\apj, {\bf 697}, 100 (2009).

\bibitem[Shukla \& Eliasson 2010]{Shukla-1} Shukla, P.K. and Eliasson, B.
Phys. Usp. {\bf 53}, 51 (2010).

\bibitem[Shukla \& Eliasson 2011]{Shukla-2} Shukla, P.K. and
Eliasson, B. Rev. Mod. Phys. {\bf 83}, 885 (2011).

\bibitem[Tajima 2014]{Dunne-3} Tajima, T. Eur. Phys. J. Sp. Top. {\bf 223(6)}, 1037 (2014).

\bibitem[Tremblay et al 2015]{tremblay} Tremblay, P.-E., Fontaine, G., Freytag, B.,
Steiner, O., \\
Ludwig, H.-G., Steffen, M., Wedemeyer, S. and Brassard, P. \apj,
{\bf 812}, 19, (2015).

\bibitem[Tsintsadze et al. 2003]{Stenflo} Tsintsadze, N.L. , Shukla, P.K. and
Stenflo, L. \ Eur. Phys. J. D. {\bf 23} 109 (2003).

\bibitem[Weinberg 1972]{Weinberg} Weinberg, S. \ 1972 \
{\it Gravitation and Cosmology} (Weley, New York).


\bibitem[Winget \& Kepler 2008]{winget} Winget, D. E.  and Kepler, S. O.
Annu. Rev. A\&A {\bf 46}, 157 (2008).

\bibitem[Yanovsky et al 2008]{Yanovski} Yanovsky, V., Chvykov, V., Kalinchenko, G.,
Rousseau, P., Planchon, T., Matsuoka, T., Maksimchuk, A., Nees,
J., Cheriaux, G. , Mourou, G.  and Krushelnick, K. \ Opt. Exp. {\bf
16} 2109 (2008).

\bibitem[Yoshida \& Mahajan 1999]{iqbal-1} Yoshida, Z.  and Mahajan, S.M.
J. Math. Phys. {\bf 40}, 5080 (1999).

\bibitem[Yoshida et al 2001]{ymois} Yoshida, Z., Mahajan, S.M., Ohsaki, S., Iqbal, M.
\& Shatashvili, N.L. Phys. Plasmas, {\bf 8(5)}, 2125 (2001).

\bibitem[Zanni et al 2007]{zanni} Zanni, C., Ferrari, A., Rosner, R., Bodo, G. and
Massaglia, S. A\&A {\bf 469}, 811 (2007).




\end{thebibliography}

%

\end{document}